# Reversible thermal strain control of oxygen vacancy ordering in an epitaxial La$_{0.5}$Sr$_{0.5}$CoO$_{3-\delta}$ film

Sampo Inkinen, Lide Yao[a)], and Sebastiaan van Dijken[a)]

*NanoSpin, Department of Applied Physics, Aalto University School of Science, P.O. Box 15100, FI-00076 Aalto, Finland*

[a)] *E-mail: lide.yao@aalto.fi and sebastiaan.van.dijken@aalto.fi*

ABSTRACT

Reversible topotactic transitions between oxygen-vacancy-ordered structures in transition metal oxides provide a promising strategy for active manipulation of material properties. While transformations between various oxygen-deficient phases have been attained in bulk ABO$_{3-\delta}$ perovskites, substrate clamping restricts the formation of distinct ordering patterns in epitaxial films. Using in-situ scanning transmission electron microscopy (STEM), we image a thermally driven reversible transition in La$_{0.5}$Sr$_{0.5}$CoO$_{3-\delta}$ films on SrTiO$_3$ from a multidomain brownmillerite (BM) structure to a uniform new phase wherein oxygen vacancies order in every third CoO$_x$ plane. Because temperature cycling is performed over a limited temperature range (25 °C – 385 °C), the oxygen deficiency parameter $\delta$ does not vary measurably. Under constant $\delta$, the topotactic transition proceeds via local reordering of oxygen vacancies driven by thermal strain. Atomic-resolution imaging reveals a two-step process whereby alternating vertically and horizontally oriented BM domains first scale in size to accommodate the strain induced by different thermal expansions of La$_{0.5}$Sr$_{0.5}$CoO$_{3-\delta}$ and SrTiO$_3$, before the new phase nucleates and quickly grows above 360 °C. Upon cooling, the film transforms back to the mixed BM phase. As the structural transition is fully reversible and $\delta$ does not change upon temperature cycling, we rule out electron-beam irradiation during STEM as the driving mechanism. Instead, our findings demonstrate that thermal strain can solely drive topotactic phase transitions in perovskite oxide films, presenting opportunities for switchable ionic devices.

Active control over the concentration or distribution of oxygen vacancies in transition metal oxides enables manipulation of their structural, magnetic, electronic transport, and optical properties via valence changes of the transition metal ions and the altering of bond lengths and angles.[1-5] In perovskite oxides, oxygen vacancies may order into distinct patterns depending on the oxygen off-stoichiometry, thereby forming a homologous series of oxygen-deficient $ABO_{(3n-1)/n}$ structures. Many members of the homologous series are accessible in bulk perovskites by topotactic oxidation or reduction reactions.[6-12] For instance, $SrCoO_{(3n-1)/n}$ with $n = 2, 3, 4, 5, 6, 7$, and 8 has been realized.[9-12] Among the oxygen-deficient perovskite phases, $ABO_{2.5}$ ($n = 2$) has been investigated most. In this phase, oxygen vacancies occupy every second $BO_x$ plane. The resultant stacking of $BO_6$ octahedra and $BO_4$ tetrahedra is known as the brownmillerite (BM) structure. BMs exhibit mixed electronic- and ionic conductivity and are relevant for gas sensors, oxygen storage devices, oxygen membranes, and cathodes in solid oxide fuel cells.[3,13-15]

In contrast to bulk samples, epitaxial films of perovskite oxides do not display the same plurality of oxygen-deficient $ABO_{(3n-1)/n}$ phases. Instead, $ABO_{3-\delta}$ films tend to favor one oxygen vacancy ordered structure, often the BM, and the perovskite phase over extended $\delta$ ranges. Epitaxial $SrCoO_{3-\delta}$ films are a good example. In this system, the $SrCoO_{2.5}$ BM phase ($n = 2$) transforms directly into the $SrCoO_3$ perovskite structure ($n = \infty$) during annealing in oxygen atmosphere at $\delta \approx 0.25$, without accessing any of the intermediate phases of the $SrCoO_{(3n-1)/n}$ homologous series.[3,15,16] Annealing in vacuum produces the reverse structural transition.[3,16] Direct phase boundaries or transformations between BM and perovskite structures are also reported for other epitaxial films, including $La_{1-x}Sr_xMnO_{3-\delta}$,[5,17,18]

SrFeO$_{3-\delta}$,[19-21] and La$_{1-x}$Sr$_x$CoO$_{3-\delta}$.[22,23] Other members of the ABO$_{(3n-1)/n}$ homologous series have been observed in some specific perovskite films. For instance, superstructures comprising a stacking of two or three CoO$_6$ octahedral layers and one CoO$_4$ tetrahedral layer, corresponding to $n = 3$ and $n = 4$, have been imaged in epitaxial LaCoO$_{3-\delta}$ films.[24-26] While several vacancy ordered structures with $2 < n < \infty$ can thus appear in perovskite films, reports on transformations between distinct oxygen-deficient phases, let alone reversible transitions, are rare.[5,27-29]

Strain imposed by the substrate can drastically affect the properties of epitaxial ABO$_{3-\delta}$ films. For instance, misfit strain tunes the oxygen deficiency ($\delta$) of the BM and perovskite phases of SrCoO$_{3-\delta}$ through a reduction of the oxygen activation energy.[30-32] Strain can also determine the orientation of oxygen vacancy ordered superstructures, as illustrated by studies on La$_{0.5}$Sr$_{0.5}$CoO$_{3-\delta}$ (LSCO).[33-35] When grown onto a LaAlO$_3$ (001) substrate with smaller lattice parameter, a lattice mismatch of $\varepsilon = +1.3\%$ compresses the LSCO film, leading to the formation of a BM structure with the oxygen-deficient CoO planes oriented parallel to the substrate surface.[33-37] Hereafter, we will refer to this phase as horizontal BM (h-BM). Growth of oxygen-deficient LSCO on SrTiO$_3$ (001) ($\varepsilon = -1.8\%$), in contrast, results in a vertical BM (v-BM) structure[34,35,38] or a mixture of h-BM and v-BM domains.[33,39] Taking into account the larger average in-plane lattice parameter of v-BM compared to h-BM, elastic arguments do explain the growth of differently oriented BM phases as an instrument to relax misfit strain through oxygen vacancy ordering.[34]

Because most experiments on topotactic phase transitions in ABO$_{3-\delta}$ films are performed by annealing in vacuum or oxygen atmosphere, the effects of strain are often masked by simultaneous changes in oxygen deficiency. Yet, strain can be a significant factor, as illustrated by strong changes in the reversibility of redox reactions in SrCoO$_{3-\delta}$ films grown onto different substrates.[32] To isolate the effects of strain on oxygen vacancy migration and ordering in perovskite films, one needs to perform experiments that allow for constant tuning of $\varepsilon$ while $\delta$ is kept constant. Here, we use in-situ scanning transmission electron microscopy (STEM) to demonstrate that thermal strain by itself can drive reversible topotactic transitions between distinct vacancy-ordered structures in an epitaxial $ABO_{3-\delta}$ film. Our results shed light on strain-structure relationships in oxygen-deficient perovskites and the atomic-scale dynamics of oxygen vacancy ordering, offering a promising mechanism for deterministic physical property control.

Cross-sectional TEM specimens of a 20-nm-thick epitaxial $La_{0.5}Sr_{0.5}CoO_{3-\delta}$ (LSCO) film on a $SrTiO_3$ (STO) substrate were prepared for high-angle annular dark-field STEM (HAADF-STEM) analysis during in-situ heating using a DENSsolutions Lightning D9+ holder (see section S1 of the supplementary material for details). In the as-grown state, the LSCO film exhibits alternating h-BM and v-BM structural domains, as illustrated by the orientation of atomic columns with bright and dark contrast in the HAADF-STEM images and clear superstructure reflections along the $x$- and $y$-axis of their fast Fourier transform (FFT) (see Fig. 1(a) and Fig. S1 of the supplementary material). The two BM structures exhibit a stacking sequence $La_{0.5}Sr_{0.5}O$ - $CoO_2$ - $La_{0.5}Sr_{0.5}O$ - $CoO_x$ along the [001] and [010] directions, respectively. Oxygen deficiency in the $CoO_x$ layer enlarges the distance between the neighboring $La_{0.5}Sr_{0.5}O$ planes (dark contrast) and it reduces the spacing of $La_{0.5}Sr_{0.5}O$ planes separated by stoichiometric $CoO_2$ (bright contrast). Using the method by Kim et al.,[40] which maps the oxygen vacancy concentration based on the local lattice parameter of the BM phase (Fig. S2 of the supplementary material), we estimate $x = 1.25 \pm 0.06$. This composition corresponds to $La_{0.5}Sr_{0.5}CoO_{2.63 \pm 0.06}$ and, thus, $\delta = 0.37 \pm 0.06$. Because the oxygen vacancy concentration is lower than in the ideal BM structure ($\delta = 0.50$), the oxygen-deficient $CoO_x$ layer comprises some $CoO_6$ octahedra or $CoO_5$ pentahedra in addition to the $CoO_4$ tetrahedra. Besides the h-BM and v-BM domains, the as grown LSCO film contains a few atomic layers with a perovskite structure near the STO substrate. Similar phase separation has been reported previously and was attributed to preferred oxygen vacancy formation near the film surface or disordering of oxygen vacancies under compressive strain near the film/substrate interface.[35,41] Small areas comprising a mixture of the two BM structures (m-BM) appear at some domain boundaries.

Next, we discuss the evolution of the LSCO film structure during in-situ heating. In the experiments, we continuously increased the temperature of the TEM specimen at a rate of 8 °C/min while recording HAADF-STEM images. Figure 1 summarizes a typical heating/cooling cycle. Here, we focus on a 30-nm-wide sample area. The selected area initially comprises two h-BM domains separated by a v-BM domain, which we colored blue and dark green for clarity (Fig. S3 of the supplementary material shows uncolored versions of the same images). Up to a temperature of 350 °C [Figs. 1(a)-1(d)], the h-BM domains grow slowly by lateral domain boundary motion at the expense of the v-BM domain. At 360 °C, a new structural phase (labeled N, purple color) nucleates at the h-BM/v-BM boundaries [Fig. 1(e)]. The appearance of extra reflections in the FFT besides the main reflections from the v-BM phase confirms the formation of a new vertically oriented superstructure. Up to 380 °C, the N phase gradually grows by replacing the original h-BM and v-BM domains and a new N domain nucleates in the center of the v-BM domain. After this, the N phase abruptly expands to the entire field-of-view at 385 °C, except for the perovskite structure (labeled P, yellow color) persisting near the STO substrate [Figs. 1(f)-1(h)]. The transition from alternating h-BM/v-BM domains to the N phase is reversible, as illustrated by HAADF-STEM imaging during cool down [Figs. 1(i)-1(o)]. The h-BM and v-BM domains nucleate at the horizontal N/perovskite boundary and grow rapidly along the vertical direction. The subsequent lateral growth of these domains is slow and far from complete after reaching room temperature in about 45 minutes [see Fig. 1(n)]. Relaxation of the domains continues at 25 °C, as confirmed by the re-establishment of alternating h-BM and v-BM domains without the N phase after storing the sample in vacuum for 15 hours [see Fig. 1(o)]. Local STEM-EELS measurements on the h-BM and v-BM domains and the N phase indicate that the heating/cooling cycle does not change the oxygen deficiency ($\delta$) of the LSCO film (Fig. S4, supplementary material). The reversible structural phase transition depicted in Fig. 1 thus demonstrates a significant re-ordering of existing oxygen vacancies as a function of temperature. Since oxygen vacancy migration is a thermally activated process, the transition from h-BM/v-BM domains to the N phase occurring at $T \geq 360$ °C is relatively fast, whereas the reverse route ensuing at lower temperatures takes more time.

We note that the fully reversible character of the experiments and the observation of constant $\delta$ rule out electron-beam irradiation as a possible driving mechanism. As previously demonstrated,[18,22,28] electron-beam irradiation during (S)TEM characterization can trigger the formation and ordering of oxygen vacancies in perovskite films. This process, however, increases $\delta$ monotonically, thereby preventing a transition back into the original structural phase. To test the influence of the electron beam further, we also tested the formation of the N phase far away from the STEM imaged area. There, the N phase formed also without any discernable structural differences.

Figures 2 compares high-resolution HAADF-STEM images of the v-BM phase at 25 °C and the N phase at 385 °C. In the v-BM structure, every second $CoO_x$ plane is oxygen deficient ($x < 2$) and the resulting lattice expansion produces stripes with dark HAADF contrast. The N structure exhibits a similar vertical stripe pattern, but now every third $CoO_x$ plane exhibits dark contrast. The period of the oxygen-vacancy-ordered structure thus changes from $2a$ to $3a$ during the thermally induced phase transition. The appearance of superstructure reflections in the FTTs of Fig. 1 at $1/3a$ and $2/3a$ along the [010] direction and the concurrent disappearance of the $1/2a$ reflections during heating to 385 °C confirm the change of the modulation period.

The N phase can be considered as the $n = 3$ member of the La$_{0.5}$Sr$_{0.5}$CoO$_{(3n-1)/n}$ homologous series. In fact, the ideal oxygen deficiency for this structure, $\delta = 0.33$, closely matches the value derived for our epitaxial LSCO film ($\delta = 0.37 \pm 0.06$). From this observation and the fact that $\delta$ remains constant during temperature cycling, we conclude that a new vertically oriented superstructure with CoO$_4$ tetrahedra in every third CoO$_x$ plane replaces the v-BM and h-BM domains with a mixture of CoO$_4$ tetrahedra, CoO$_5$ pentahedra, and CoO$_6$ octahedra in every second CoO$_x$ plane. Figures 2(e) and 2(f) show structural models for the v-BM and N phase together with in-plane lattice spacing maps corresponding to the STEM images of Figs. 2(c)-(d). The average pseudo-cubic lattice parameter along [010] measured far away from the LSCO/STO interface is $a_{pc} = 3.98 \pm 0.01$ Å for the v-BM structure and $a_{pc} = 3.89 \pm 0.01$ Å for the N phase.

Because the concentration of oxygen vacancies remains constant during the in-situ STEM experiments, strain is the driving force behind the thermally induced structural phase transitions. In our non-stoichiometric LSCO film, the formation of alternating h-BM and v-BM domains reduces the elastic energy at room temperature.[33,39] The lattice mismatch between the h-BM phase ($a_{pc} = 3.86 \pm 0.01$ Å, measured far away from STO) and STO ($a = 3.905$ Å) is $-1.2\%$, whereas the lattice mismatch between v-BM domains ($a_{pc} = 3.98 \pm 0.01$ Å on average along [010]) and STO is $+1.9\%$. Thus based on elastic arguments,[34] one would expect the more heavily strained v-BM domains to be smaller. Analyzing a large set of room-temperature STEM images, we find that this is indeed the case with the h-BM and v-BM domains having an average width of 45 nm and 20 nm, respectively. Upon heating the LSCO/STO sample, the strain state of the LSCO film changes because the linear thermal expansion coefficient of LSCO ($1.79 \times 10^{-5}$ K$^{-1}$)[42] is larger than that of STO ($1.08 \times 10^{-5}$ K$^{-1}$).[43] Consequently, the compressive strain in the v-BM domains increases, while the tensile strain in the h-BM domains diminishes. Because of opposite changes in the elastic energy of the two domains, the h-BM phase grows slowly at the expense of the v-BM phase as the temperature increases, in agreement with the HAADF-STEM images of Figs. 1(a)-1(d). The in-plane strain maps and line profiles of Figs. 3(a)-(b), calculated by geometric phase analysis (GPA) of the images in Fig. 1, illustrate the evolution of lattice strain during temperature cycling. The compressive strain in the v-BM domain increases from +1.9% at room temperature to about +2.3% at 350 °C [(Fig. 3(c)], which shrinks the domain gradually [(Fig. 3(d)]. Further heating enhances the strain in the v-BM domain to an unsustainable level, triggering the nucleation of the N phase. The N phase nucleates at h-BM/v-BM boundaries and within the v-BM domain. The nucleation locations do demonstrate that relaxation of the thermally enhanced compressive strain in the v-BM domain indeed acts as the driving force behind the creation of N domains. Once formed, the N domains grow quickly at the expense of both the v-BM and h-BM phases until their replacement is complete at 385 °C [see Figs. 1(h) and Fig. 3(d)]. The disappearance of the h-BM domains at high temperature in favor of the N phase suggests that the latter is strained less on STO. While this conclusion is supported in part by the experimental data on in-plane strain along the [010] direction (–0.8% tensile strain in the N domain at 380 °C versus –1.0% strain in the h-BM domain at the same temperature, see Fig. 3(c)), the lattice mismatch along [100] (i.e. the direction of the electron beam in our in-situ STEM experiments) should be considered in a full strain analysis. Finally, we note that the growth of the center and right N domains (see data for 380 °C in Fig. 3(b)) produces a stacking fault with an ill-defined GPA strain state upon domain coalescence at 385 °C. In the stacking fault, the period of the oxygen-vacancy-ordered structure is $2a$ instead of $3a$. Upon cooling, a v-BM domain grows from the stacking fault at T < 280 °C to relax the tensile strain of the N phase. Simultaneously, h-BM domains reform and the growth of BM domains ultimately restores an alternating h-BM/v-BM domain pattern at room temperature. For a full GPA data set including out-of-plane strain maps see Fig. S5 of the supplementary material.

Figure 4 visualizes the atomic scale dynamics of oxygen vacancy reordering during reversible transitions between the v-BM and N structures. Here, we focus on a LSCO film area located in a v-BM domain at room temperature (marked green in Fig. 4(a)). Upon heating the period of the oxygen-deficient $CoO_x$ planes changes from $2a$ to $3a$ [Fig. 4(b)]. Figure 4(c) shows intensity profiles of HAADF-STEM images as a function of temperature to elucidate this process. Herein, deep intensity minima (dark contrast) correspond to oxygen-deficient $CoO_x$ planes, shallow minima mark the location of stoichiometric $CoO_2$ planes, and intensity maxima depict $La_{0.5}Sr_{0.5}O$ planes. Starting at 25 °C, oxygen vacancies occupy every second $CoO_x$ plane (v-BM phase). During the transition to the N structure at elevated temperatures, the position of every third oxygen-deficient $CoO_x$ plane remains fixed (marked by dashed red lines in Fig. 4(c)). Vacancies in the other oxygen-deficient $CoO_x$ planes migrate to the oxygen stoichiometric $CoO_2$ plane that separate them initially (see black dashed lines and arrows in Fig. 4(c)). As a result, the modulation period of the vertical oxygen-vacancy-ordered structure changes from $2a$ to $3a$. The oxygen-vacancy-driven transformations in the HAADF-STEM intensity profiles [Fig. 4(c)] and in-plane lattice spacing maps [Fig. 4(b)] are relatively fast, demonstrating completion of the v-BM to N phase transition in a narrow 360 °C – 385 °C temperature window. Upon cooling to room temperature, the reverse migration process ensues. Now, the position of every second oxygen-deficient $CoO_x$ plane remains fixed (red dashed line in Fig. 4(c)), whereas the others split into two (black dashed lines and arrows in Fig. 4(c)). Because the N to v-BM phase transition occurs at lower temperature, oxygen vacancy migration is less efficient. As a result, the transition takes more time and a larger degree of oxygen vacancy disorder develops during the structural transformation. The lattice spacing maps obtained at 280 °C and 225 °C [Fig. 4(b)] illustrate the latter point particularly well. In the selected LSCO film area of Fig. 4, the v-BM phase is fully reestablished at 200 °C. Interestingly, since the two $CoO_x$ planes that combine into one during heating and the one $CoO_x$ plane that splits into two during cooling are different in location, the v-BM superstructure undergoes a lateral shift by lattice parameter *a* during the full heating/cooling cycle. A similar analysis of oxygen vacancy reordering during the nucleation of the N phase at a h-BM/v-BM domain boundary is presented in Fig. S6 of the supplementary material. Atomic-resolution HAADF-STEM images of the h-BM/v-BM boundary area reveal that the N phase nucleates on the v-BM side of the boundary, after which it quickly expands into both BM domains.

In summary, using in-situ sample heating and STEM imaging, we demonstrate reversible topotactic transitions between two distinct oxygen-vacancy-ordered structures in an epitaxial $La_{0.5}Sr_{0.5}CoO_{3-\delta}$ film on $SrTiO_3$ (001). In contrast to studies where the oxygen deficiency parameter $\delta$ of a perovskite film is varied to induce a structural phase transition (oxidation/reduction reactions, electron beam irradiation, electric-field control),[3-5,15-23,27-32] $\delta = 0.37 \pm 0.06$ is kept constant in our experiments. Under fixed oxygen off-stoichiometry, we find that thermal strain controls the occurrence of topotactic phase transitions. In our material system, a pattern of alternating h-BM domains (tensile strain) and v-BM domains (compressive strain) is energetically favorable at room temperature. The elastic energy of the LSCO film changes upon heating because its thermal expansion coefficient is larger than that of the STO substrate. Initially, this leads to the growth of h-BM domains at the expense of v-BM domains, but at 360 °C a new oxygen-vacancy-ordered structure emerges as the lowest energy state under thermal strain. In the new phase, which covers the entire LSCO film at 385 °C, oxygen vacancies order into every third $CoO_x$ plane rather than every second in the BM phase. Microscopically, a redistribution of oxygen vacancies whereby some $CoO_x$ planes remain fixed and others combine into one explains the change of the modulation period from 2*a* to 3*a*. Upon cooling, the reverse topotactic phase transition ensues on a longer timescale because of reduced oxygen vacancy mobility at lower temperature.

Our results point to interesting perspectives. First, the presented in-situ STEM experiments demonstrate a great sensitivity of oxygen-vacancy-ordered phases to small changes in lattice strain. Together with the strong dependence on the oxygen deficiency parameter, this opens up a playground for the engineering of distinct topotactic transitions in closed epitaxial systems with constant oxygen off-stoichiometry. Second, thermal strain control of oxygen vacancy ordering in perovskite transition metal oxide films opens the door to reversible property manipulation, including magnetism, ferroelectricity, optical reflection/transmission, and electronic transport. Compared to crystalline-amorphous transitions in phase-change materials such as chalcogenide alloys that mainly alter their electrical resistivity,[44,45] this broadens the application potential of thermally induced structural phase transitions.

See the supplementary material for information on the growth of LSCO films on STO, the preparation of cross-sectional specimen, the in-situ heating STEM measurements, and supporting experimental data.

This work was supported by the Academy of Finland (Grant Nos. 293929, 304291, 319218, and 316857). TEM analysis was conducted at the Aalto University OtaNano-Nanomicroscopy Center (Aalto-NMC).

**Figures and captions**

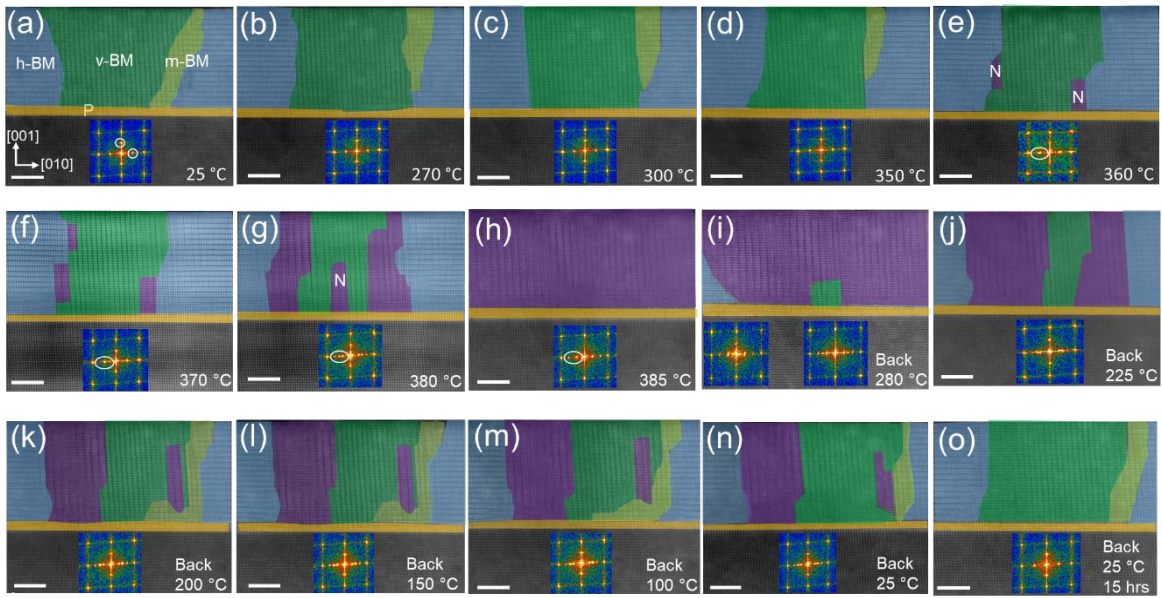

**FIG. 1.** Evolution of oxygen-vacancy-ordered domains during temperature cycling. [(a)-(o)] HAADF-STEM images of a 20-nm-thick epitaxial $La_{0.5}Sr_{0.5}CoO_{3-\delta}$ film on a $SrTiO_3$ substrate recorded during in-situ heating and cooling. For clarity the structures are labeled and colored: perovskite (P, yellow), horizontal brownmillerite (h-BM, blue), vertical brownmillerite (v-BM, dark green), mixed h-BM and v-BM (m-BM, light green), and new phase (N, purple). The scale bars correspond to 5 nm. Fast Fourier transform (FTT) patterns of the HAADF-STEM images are shown as insets. White ellipses in the FTT patterns mark superstructure reflections of the v-BM phase (period $2a$) and N phase (period $3a$).

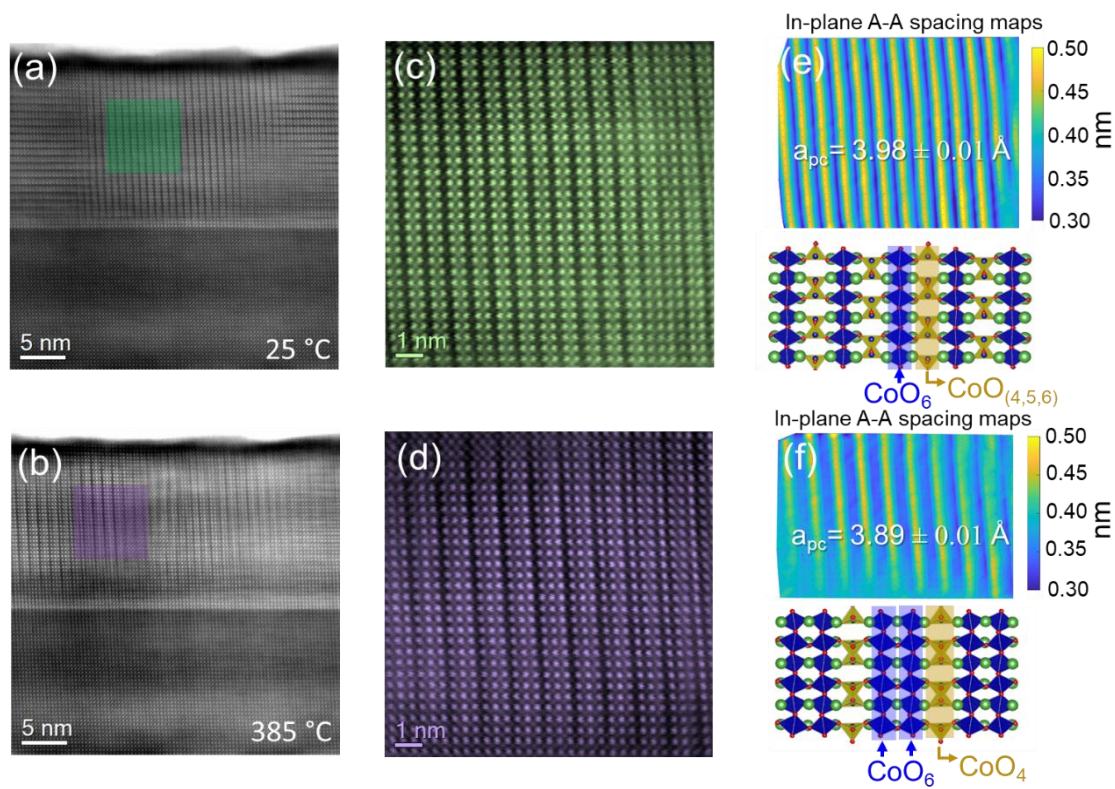

**FIG. 2.** [(a)-(d)] HAADF-STEM images and high-magnification close ups of a v-BM domain recorded at 25 °C [(a) and (c)] and an N domain recorded at 385 °C [(b) and (d)]. Vertical lines with dark HAADF contrast indicate oxygen-deficient $CoO_x$ planes with an expanded lattice parameter. [(e) and (f)] In-plane lattice spacing maps and structural models of (e) the v-BM phase and (f) the N phase.

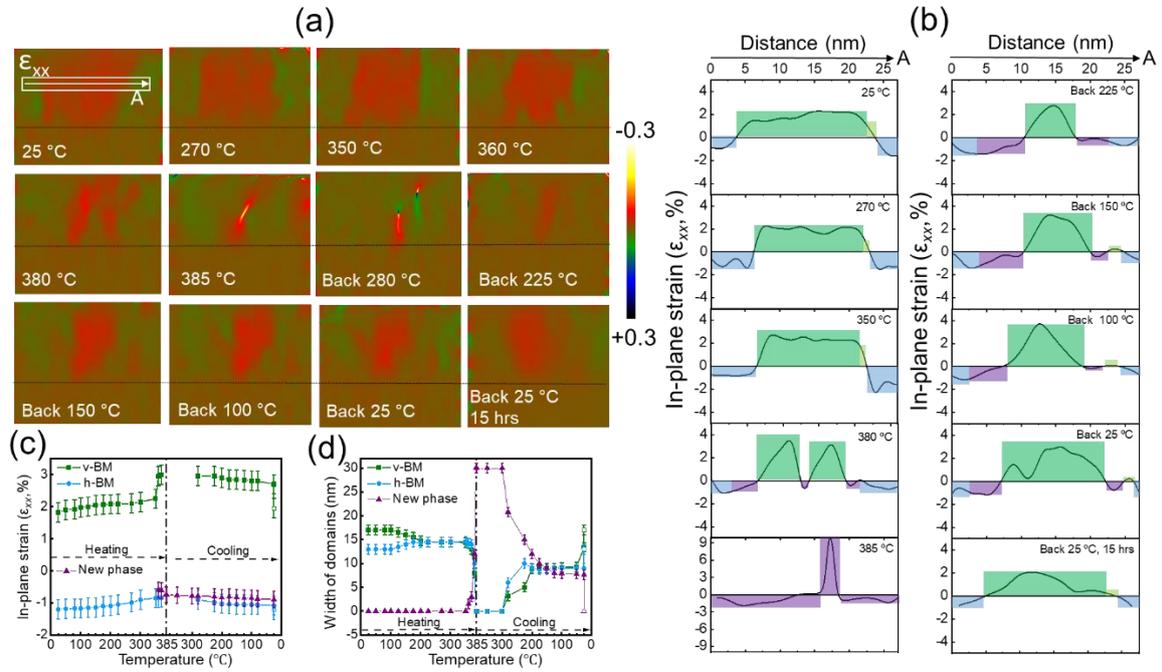

**FIG. 3.** (a) In-plane strain maps calculated from selected HAADF-STEM images in Fig. 1 (a complete set of in-plane and out-of-plane strain maps is shown in Fig. S5 of the supplementary material). The dashed lines indicate the LSCO/STO interface. (b) In-plane strain line profiles extracted from the maps in (a). The data are collected along the arrow A and averaged over the area marked by a white box. Positive values of $\varepsilon_{xx}$ represent a compressive in-plane strain along [010] (lattice parameters of LSCO phase > STO), whereas negative values of $\varepsilon_{xx}$ indicate tensile strain. In (b) the same colors are used as in Fig. 1 to mark the structural domains. (c) Evolution of in-plane strain as a function of temperature in the different domains. (d) Variation of the domain widths during heating and cooling.

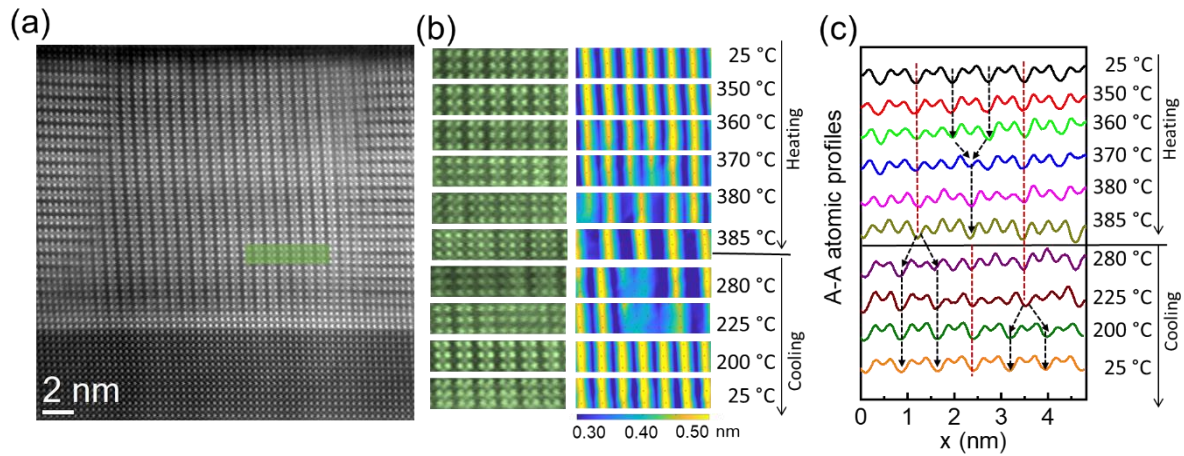

**FIG. 4.** Atomic-scale analysis of oxygen-vacancy migration dynamics during reversible transitions between the v-BM and N phases. (a) HAADF-STEM image of the initial LSCO film structure with the area under study marked in green. (b) Close ups of the green area recorded at different temperatures during heating and cooling. Corresponding in-plane lattice spacing maps are shown on the right. (c) HAADF-STEM intensity profiles extracted from the images in (b). Red dashed lines mark the oxygen-deficient $CoO_x$ planes whose location remain fixed during heating/cooling. Black dashed lines and arrows indicate how two $CoO_x$ planes combine into one (heating) or how one $CoO_x$ plane splits into two (cooling).

# Supplementary material

# Reversible Thermal Strain Control of Oxygen Vacancy Ordering in an Epitaxial La$_{0.5}$Sr$_{0.5}$CoO$_{3-\delta}$ Film

Sampo Inkinen, Lide Yao[a], and Sebastiaan van Dijken[a]

*NanoSpin, Department of Applied Physics, Aalto University School of Science, P.O. Box 15100, FI-00076 Aalto, Finland*

[a] *E-mail: lide.yao@aalto.fi and sebastiaan.van.dijken@aalto.fi*

## S1: Sample fabrication and in-situ heating STEM experiments

A 20-nm-thick La$_{0.5}$Sr$_{0.5}$CoO$_{3-\delta}$ film was epitaxially grown on a single-crystal SrTiO$_3$ substrate using pulsed laser deposition (PLD). Before film growth, the substrate was etched in buffered hydrofluoric acid for 30 seconds and annealed in oxygen atmosphere at 950 °C for 1 hour. This preparation process resulted in a TiO$_2$-terminated surface. We operated the PLD system using a 2.5 J cm$^{-2}$ laser fluence and a 4 Hz pulse repetition rate. The La$_{0.5}$Sr$_{0.5}$CoO$_{3-\delta}$ film was deposited at 800 °C in an oxygen partial pressure of 0.2 mbar. After growth, we cooled the sample to room temperature in about 50 minutes using the same oxygen environment.

Cross-sectional STEM lamella were prepared by focused ion beam using Ga$^+$ ions accelerated at 30 kV in a FEI Helios Nanolab 600 system. To reduce charging and ion beam damage during FIB processing, we deposited a 30-nm-thick Pt/Ir conductive layer on top of the LSCO film using a sputtering system (Leica EM ACE600) before loading the sample into the FIB chamber. Next, a 1.5-μm-thick Pt structure with dimension 10 μm × 2 μm was grown using ion-beam-induced deposition inside the FIB. After that, we defined a lamella by milling 6-μm-deep grooves next to the Pt-capped area. The lamella was lifted out of the sample using an OmniProbe manipulator and transferred onto a DENSsolutions heating nanochip. Finally, the cross-sectional lamella was fine-polished to about 50 nm in thickness by Ar ion milling and the nanochip was mounted onto a DENSsolutions lightning D9+ holder for in-situ STEM characterization.

HAADF-STEM imaging (see Fig. S1) and local EELS analyses (see Fig. S4) were carried out using a double Cs corrected JEOL FS2200 system operated at 200 kV. During the heating/cooling cycles we varied the sample temperature at a rate of 8 °C/min. Thermal and epitaxial strains of the film were evaluated by geometric phase analysis of HAADF-STEM images using the FRWRTools plugin[1,2] implemented in DigitalMicrograph (Gatan, Inc.). For lattice analysis, the coordinates of the atomic columns were extracted using a 2D Gaussian fitting script in Matlab and the lattice spacings were calculated by the Peak Pair algorithm.[3]

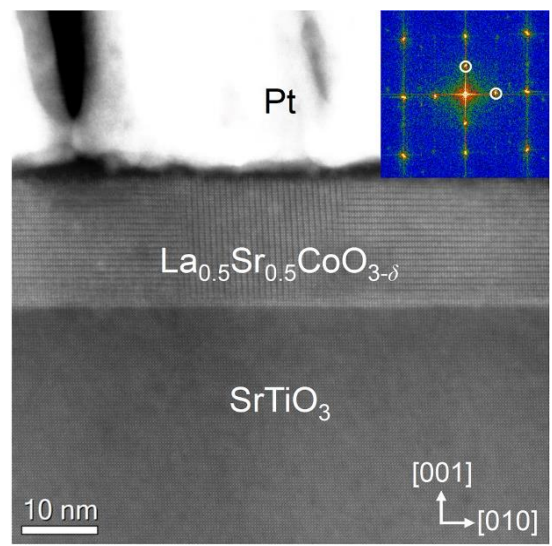

**FIG. S1**. Cross-sectional HAADF-STEM image of the as-grown 20-nm-thick $La_{0.5}Sr_{0.5}CoO_{3-\delta}$ film on a $SrTiO_3$ substrate. The film consist of alternating h-BM and v-BM domains. The inset show a FTT pattern of the HAADF-STEM image. The circles in the FTT pattern mark the superstructure reflections of the two BM phases.

## S2. Oxygen deficiency in the $La_{0.5}Sr_{0.5}CoO_{3-\delta}$ film

Oxygen vacancies in perovskite oxides ($ABO_{3-\delta}$) can order into the BM structure. The BM structure exhibits a stacking sequence $AO$-$BO_2$-$AO$-$BO_x$ along the modulation direction (i.e. [001] for h-BM and [010] for v-BM in Figure S1). Oxygen deficiency in the $BO_x$ layer enlarges the distance between neighboring AO planes. It has be shown previously that the chemical expansion of the lattice scales approximately linearly with decreasing $x$.[4] Here, we use this relationship to estimate the values of $x$ and $\delta$ ($\delta = (2 - x)/2$) in our $La_{0.5}Sr_{0.5}CoO_{3-\delta}$ film from the lattice parameter of the h-BM structure. In order to limit the effect of epitaxial strain on the BM lattice, we extracted the out-of-plane lattice parameter of the h-BM phase far away from the $SrTiO_3$ substrate. The chemically expanded lattice spacing between neighboring $La_{0.5}Sr_{0.5}O$ planes in the h-BM phase (dark contrast in STEM-HAADF images) is measured as $4.35 \pm 0.08$ Å. Using reference points for the nominal perovskite phase ($x = 2$, $c = 3.836$ Å) and the nominal BM phase ($x = 1$, $c = 4.53$ Å),[4] we estimate $x = 1.25 \pm 0.06$ and $\delta = 0.37 \pm 0.06$ for our $La_{0.5}Sr_{0.5}CoO_{3-\delta}$ film (see Fig. S2). STEM-EELS analysis indicates that the concentration of oxygen vacancies remains constant during temperature cycling in the in-situ STEM experiments (see Fig. S4).

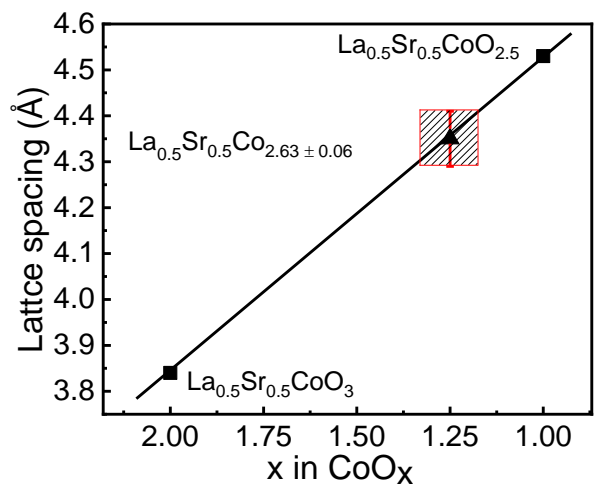

**FIG. S2.** Out-of-plane lattice spacing as a function of oxygen content $x$ in the $CoO_x$ planes of $La_{0.5}Sr_{0.5}CoO_{3-\delta}$. As reference points, we plot the known lattice spacing of the nominal perovskite $La_{0.5}Sr_{0.5}CoO_3$ and nominal BM $La_{0.5}Sr_{0.5}CoO_{2.5}$. Assuming a linear relationship between lattice spacing and oxygen content $x$, we estimate the oxygen content of our $La_{0.5}Sr_{0.5}CoO_{3-\delta}$ film by placing the measured value of the chemically expanded lattice spacing

in the h-BM phase ($c = 4.35 \pm 0.08$ Å) onto the solid line. This gives $x = 1.25 \pm 0.06$ and $\delta = 0.37 \pm 0.06$.

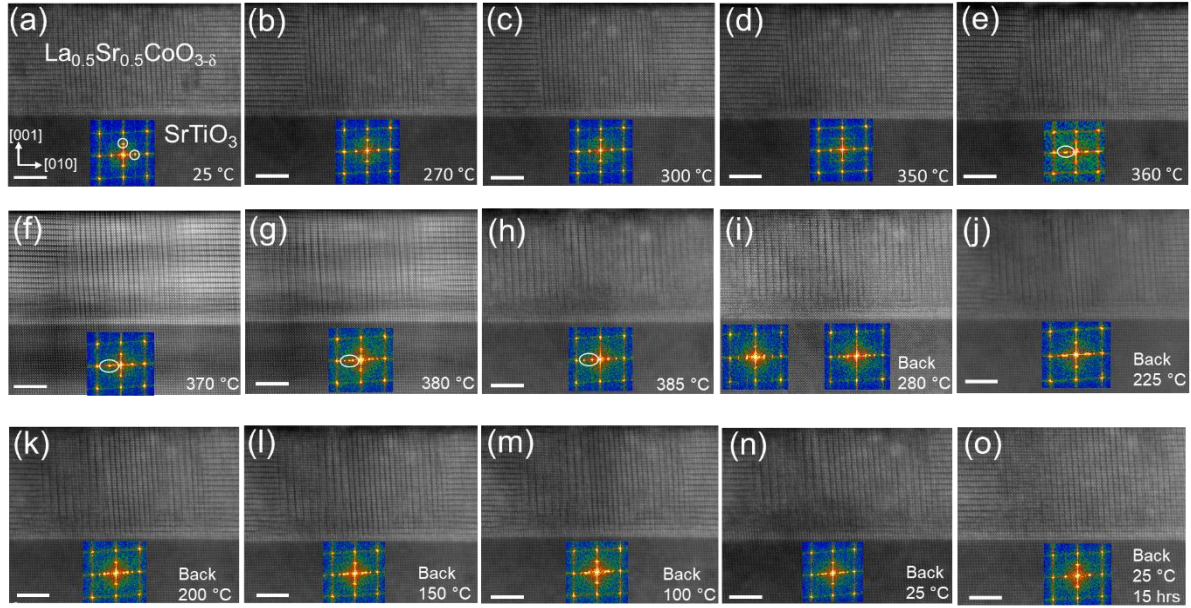

**FIG. S3**. Raw uncolored HAADF-STEM images depicting the temperature evolution of structural domains in the 20-nm-thick epitaxial $La_{0.5}Sr_{0.5}CoO_{3-\delta}$ film on a SrTiO3 substrate. The image sequence corresponds to the one shown in Fig. 1 of the main manuscript.

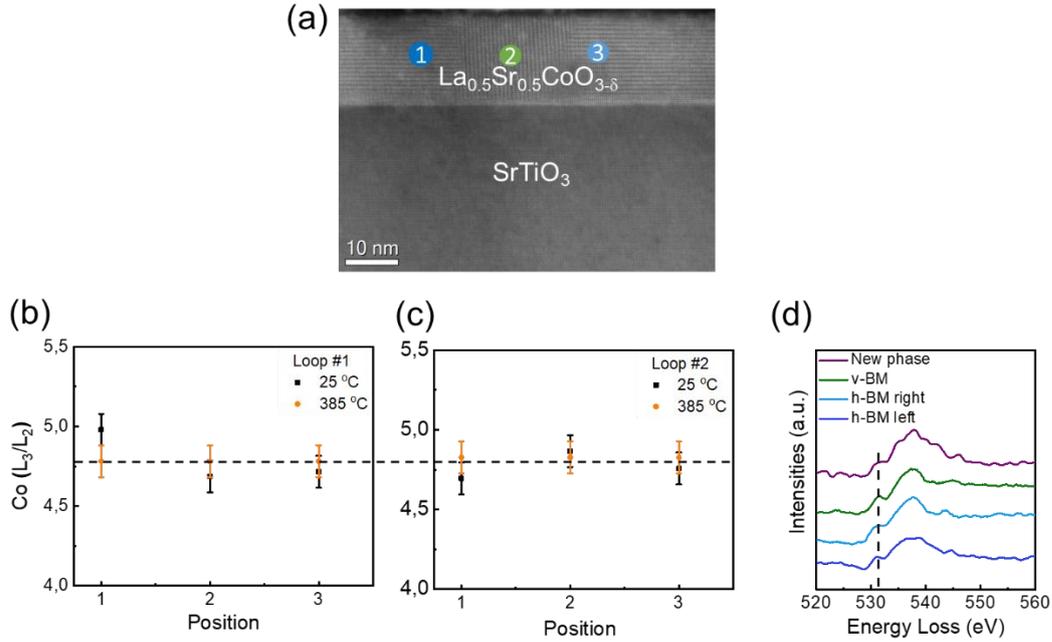

**FIG. S4**. STEM-EELS analysis of the oxygen content at different locations of the $La_{0.5}Sr_{0.5}CoO_{3-\delta}$ film. (a) EELS probing positions. At room temperature, probing positions 1 and 3 are located in h-BM domains, whereas probing position 2 is located in a v-BM domain. At 385 °C, all positions are located in the new N phase. [(b) and (c)] Co $L_3/L_2$ white line ratios at different positions calculated by the Pearson's method.[5] The black and orange data points are measured at 25 °C and 385 °C. Panels (b) and (c) present STEM-EELS results obtained during the first and second heating cycle, respectively. (d) O K-edge EELS spectra recorded on the h-BM, V-BM and N structures. The absence of discernable differences in the Co $L_3/L_2$ white line ratios [(b) and (c)] and the position and intensity of the O K-edge pre-peak (dashed line in (d)) indicates that the heating/cooling cycle does not change the oxygen deficiency of the $La_{0.5}Sr_{0.5}CoO_{3-\delta}$ film.

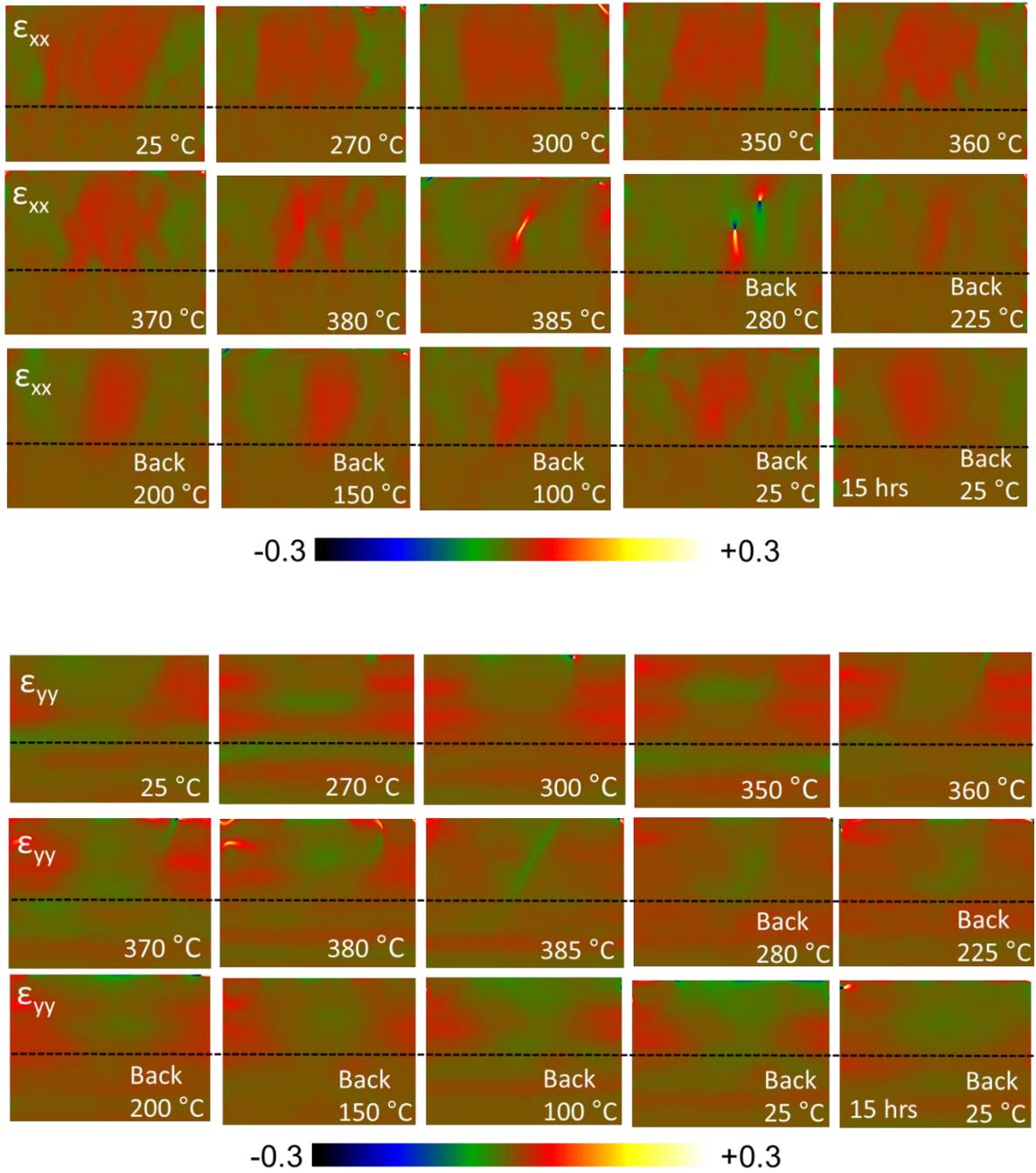

**FIG. S5**. In-plane ($\varepsilon_{xx}$) and out-of-plane ($\varepsilon_{yy}$) strain maps calculated from the HAADF-STEM images in Fig. 1 of the main manuscript. The dashed lines indicate the LSCO/STO interface.

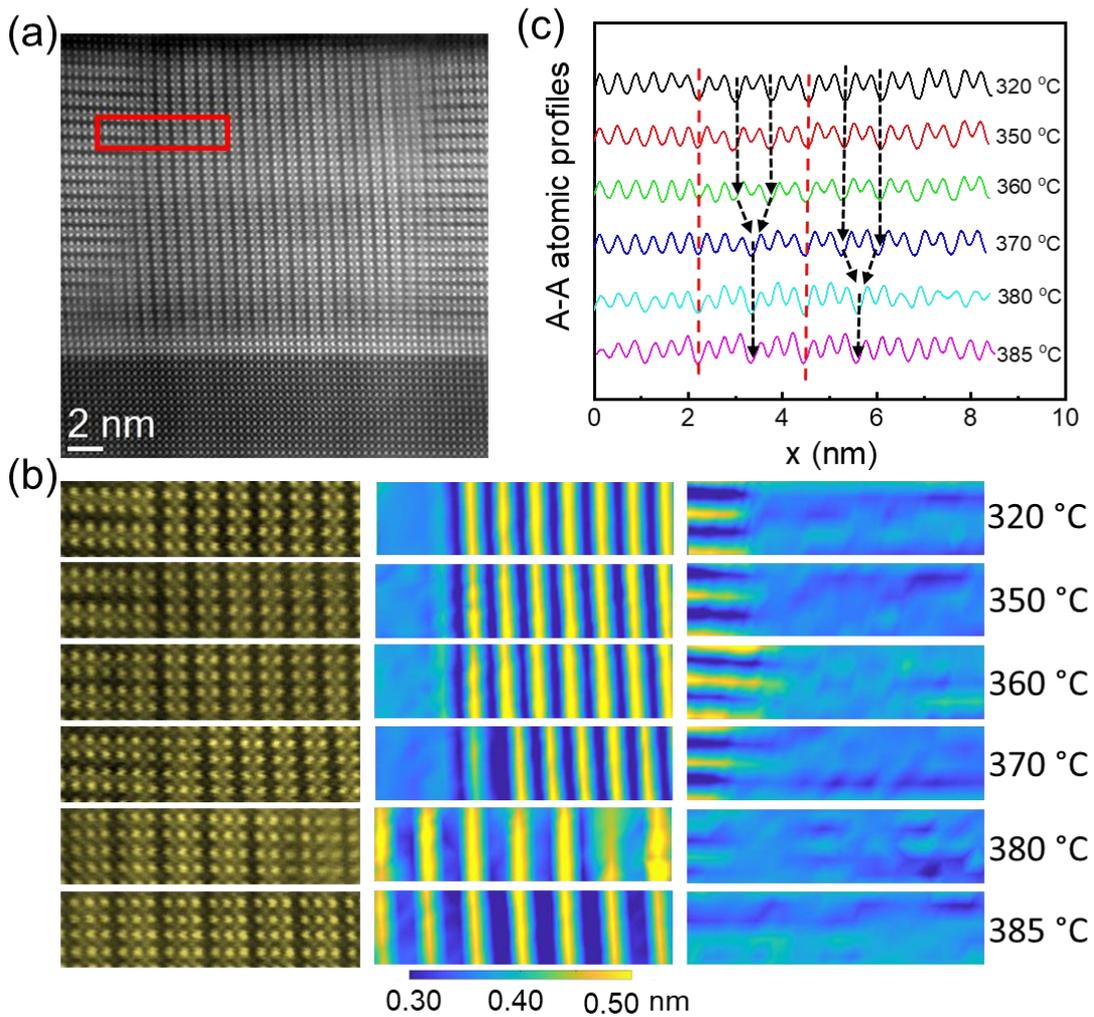

**FIG. S6**. Atomic-scale analysis of oxygen-vacancy migration dynamics during the nucleation of the N phase at a h-BM/v-BM domain boundary. (a) HAADF-STEM image of the initial LSCO film structure with the area under study marked by a red box. (b) Close ups of the boxed area in (a) recorded at different temperatures during heating (left column). Corresponding in-plane lattice spacing maps and out-of-plane lattice spacing maps are shown in the middle and right column. (c) HAADF-STEM intensity profiles extracted from the images in (b). Black dashed lines and arrows indicate how two $CoO_x$ planes in the v-BM phase combine into one in the N phase during heating. The N phase nucleates at 370 °C on the v-BM side of the boundary. Hereafter, it quickly expands into both domains.